# MonArch: Network Slice Monitoring Architecture for Cloud Native 5G Deployments

Niloy Saha*, Nashid Shahriar†, Raouf Boutaba* and Aladdin Saleh‡
{n6saha, rboutaba}@uwaterloo.ca, nashid.shahriar@uregina.ca, aladdin.saleh@rci.rogers.com
*University of Waterloo, Canada, †University of Regina, Canada, ‡Rogers Communications, Canada Inc.

***Abstract*—Automated decision making algorithms are expected to play a key role in management and orchestration of network slices in 5G and beyond networks. State-of-the-art algorithms for automated orchestration and management tend to rely on data-driven methods which require a timely and accurate view of the network. Accurately monitoring an end-to-end (E2E) network slice requires a scalable monitoring architecture that facilitates collection and correlation of data from various network segments comprising the slice. The state-of-the-art on 5G monitoring mostly focuses on scalability, falling short in providing explicit support for network slicing and computing network slice key performance indicators (KPIs). To fill this gap, in this paper, we present *MonArch*, a scalable monitoring architecture for 5G, which focuses on network slice monitoring, slice KPI computation, and an application programming interface (API) for specifying slice monitoring requests. We validate the proposed architecture by implementing MonArch on a 5G testbed, and demonstrate its capability to compute a network slice KPI (e.g., slice throughput). Our evaluations show that MonArch does not significantly increase data ingestion time when scaling the number of slices and that a 5-second monitoring interval offers a good balance between monitoring overhead and accuracy.**

***Index Terms*—Network Slicing, KPI, Monitoring, 5G Network**

## I. Introduction

Network slicing enables the creation of multiple isolated virtual networks for different services on a common physical infrastructure, but it also increases network complexity and requires the use of automated orchestration and management using Artificial Intelligence/Machine Learning (AI/ML) techniques [1]. State-of-the-art algorithms for automated orchestration and management tend to rely on data-driven methods to extract large volumes of data from the network efficiently. However, monitoring an E2E slice involves monitoring different types of resources such as radio, network, computing, and storage from multiple network segments, which poses a significant challenge in obtaining a timely and accurate view of the 5G network. Existing monitoring tools are focused on gathering information from a particular segment of the network and lack a cohesive view of an entire E2E network slice.

A key aspect in developing an E2E monitoring solution for 5G network slicing is to develop a scalable architecture to collect and correlate monitoring data from different segments of the network to compute slice KPIs. Additionally, the next-generation 5G core (5GC) is increasingly relying on cloud-native technologies (such as containers) with the 3rd Generation Partnership Project (3GPP) adopting a service-based architecture for the 5GC in Release 15 and beyond [2]. The adoption of cloud-native technologies allows supporting multiple distributed network functions (NFs) per network slice, but this also requires network orchestration and management to leverage container orchestration tools such as OpenShift and Kubernetes. This introduces new challenges for monitoring, since traditional monitoring tools cannot easily observe ephemeral cloud-native elements such as containers. Therefore, there is a need for a novel monitoring architecture for 5G and beyond networks, which provides seamless integration with cloud-native orchestration tools.

The recent literature [3]–[6] on a scalable monitoring architecture for 5G and beyond networks has focused on addressing the scalability aspect, but falls short in providing support for network slicing, concrete implementation for computing slice KPIs, and APIs for specifying monitoring requests at the network slice level. Some works including [3], [4] have no support for network slicing, while others such as [5], [6] mention it in an abstract architectural standpoint without providing any details on how actual slice monitoring and computation of KPIs may be achieved. To effectively evaluate the quality of service (QoS) requirements for 5G services, standardization bodies such as 3GPP have defined several slice KPIs such as throughput, latency, and reliability [7]. These KPIs are often composite, requiring the collection and correlation of multiple infrastructure and NF related metrics from different network segments. A monitoring architecture suitable for 5G should facilitate the collection and correlation of such metrics and enable the computation of network slice KPIs. Additionally, it should include abstractions that allow applications to specify monitoring requests at a high-level and focus on their business logic.

In this paper, we present MonArch, a comprehensive and scalable monitoring architecture for 5G and beyond networks. Our main focus is on monitoring network slices, computing slice KPIs, and providing an API for specifying monitoring requests. Our contributions include:

- A scalable monitoring architecture that can monitor at different levels such as network slices, network functions, and infrastructure.
- A northbound API that allows for specifying monitoring requests at various levels, including slices and network functions.
- A prototype implementation of the proposed monitoring system, validated through a concrete use case of monitoring network slice throughput KPI, which is publicly available on GitHub [8].
- An evaluation of the proposed architecture using a 5G testbed, showcasing the impact of scaling the number of slices and varying the monitoring interval.

The rest of the paper is organized as follows. In Section II, we survey the state-of-the-art. Section III presents the proposed monitoring architecture and Section IV discusses its implementation, focusing an a concrete example of network slice throughput KPI computation. Evaluation results are presented in Section V, and in Section VI, we conclude the paper and discuss future work.

## II. Literature Survey

In recent literature, there have been a number of works that have focused on scalable monitoring for 5G networks. One such example is the work of Perez et al. [3], in which they present a scalable monitoring architecture for a multi-site 5G platform. The proposed architecture consists of a two-level hierarchy of publish-subscribe brokers, both intra-site and inter-site, along with infrastructure-specific agents for extraction and translation of metrics from heterogeneous infrastructure components. While the authors in [3] address scalability by including a distributed Apache Kafka broker, they do not discuss important aspects of a 5G monitoring system such as end-to-end monitoring and network slicing.

Another example of 5G monitoring literature is the work of Giannopoulos et al. [4], in which they present a monitoring framework for 5G systems that consists of a suite of open-source monitoring tools. The authors focus on monitoring different levels of the 5G system, including the network slice, virtualized NF, and infrastructure level, using separate instances of monitoring tools such as Netdata and Prometheus. They demonstrate how these tools can be used to collect system and network level metrics by deploying their proposed framework on a testbed consisting of a physical gNB and the open-source Open5GS 5G SA core. However, they do not discuss how to capture per-slice KPI measurements.

Beltrami et al. [5] propose an architecture for monitoring end-to-end network slices across multiple heterogeneous domains. The focus of this work is on horizontal and vertical elasticity of the monitoring architecture by on-demand instantiation/deletion of monitoring agents in the NFs comprising a slice. This is achieved using two components: a) slice measurements aggregator(s) and b) adapters, which are deployed and managed per slice. The aggregator function is responsible for merging and formatting streams of monitoring data, while the adapters provide technology-specific southbound interfaces to monitoring solutions such as SNMP. The authors propose an engine controller component for tuning the configuration of the monitoring components (e.g., monitoring frequency), however they do not provide any quantitative analysis or results related to resource consumption of the proposed architecture.

In [6], Mekki et al. present a scalable monitoring architecture with a focus on network slices, including collecting metrics from different network slice segments. This is achieved by deploying domain (e.g., RAN and edge/core) and slice-specific collection agents, instantiated per slice, which collect and aggregate monitoring data per slice. Slice identification is achieved using a custom protocol that encapsulates monitoring data with a header containing slice identifiers. While this architecture introduces several useful abstractions, the paper falls short in describing how slice-level aggregations are done as well as how domain orchestrators compute KPIs from monitoring data of NFs in a slice.

Vasilakos et al. [9] proposed ElasticSDK, a software development kit (SDK) which provides abstractions for the development and chaining of monitoring applications. The proposed SDK leverages a distributed ElasticSearch database to provide scalability and the capability to perform aggregation queries on the data. The authors utilized a custom agent deployed on top of the FlexRAN controller to collect monitoring data and showed how ElasticSDK can be used to establish a monitoring pipeline with control plane applications reading from the database, computing processed values, and subsequently writing back to the database. The authors discuss the need for a northbound API for monitoring requests, but no details are presented regarding how filtering and aggregation can be realized based on criteria such as network slices involving different segments of the network.

TABLE I: Comparison of existing literature and proposed architecture (MonArch). Legends in the table represent supported (✓), not supported (✗), and partial/limited support (◆).

| Ref. | Data Collection | Monitoring API | E2E Monitoring | Slice KPI | Scaling/ Elasticity |
|---|---|---|---|---|---|
| [3] | Monitoring agent per component. | ✗ | ✗ | ✗ | ✓ |
| [4] | Custom NetData plugin. | ✗ | ✓ | ✗ | ✗ |
| [5] | Monitoring agent per infrastructure component. | ✗ | ✗ | ◆ | ✓ |
| [6] | FlexRAN and Prometheus exporters. | ◆ | ✗ | ◆ | ✓ |
| [9] | FlexRAN API. | ◆ | ✗ | ✗ | ✓ |
| MonArch | Prometheus exporters and FileBeats. | ✓ | ✓ (through plug-ins) | ✓ | ✓ |

**Synthesis:** The existing literature on 5G monitoring has primarily focused on scalable monitoring architecture, but lacks support for network slicing and implementation using a 5G testbed. Our work fills this gap by presenting MonArch, a scalable monitoring architecture for 5G and beyond networks, which has a specific focus on monitoring at the network slice level, slice KPI computation, and an API for specifying monitoring requests. We have also implemented and evaluated the proposed architecture using a cloud-native 5G testbed, and provided an empirical analysis on its ability to calculate 5G network slice KPIs. Additionally, our framework introduces a northbound API that allows for the conversion of high-level monitoring requests from applications into low-level monitoring primitives, which are then autonomously deployed in the 5G network. To the best of our knowledge, this is the first work to demonstrate and evaluate a scalable monitoring architecture for 5G networks that includes these features.

## III. MonArch

In this section, we outline the conceptual architecture of MonArch, detailing the flow of a monitoring request through the system and discussing the Northbound API for monitoring.

### A. Conceptual Architecture

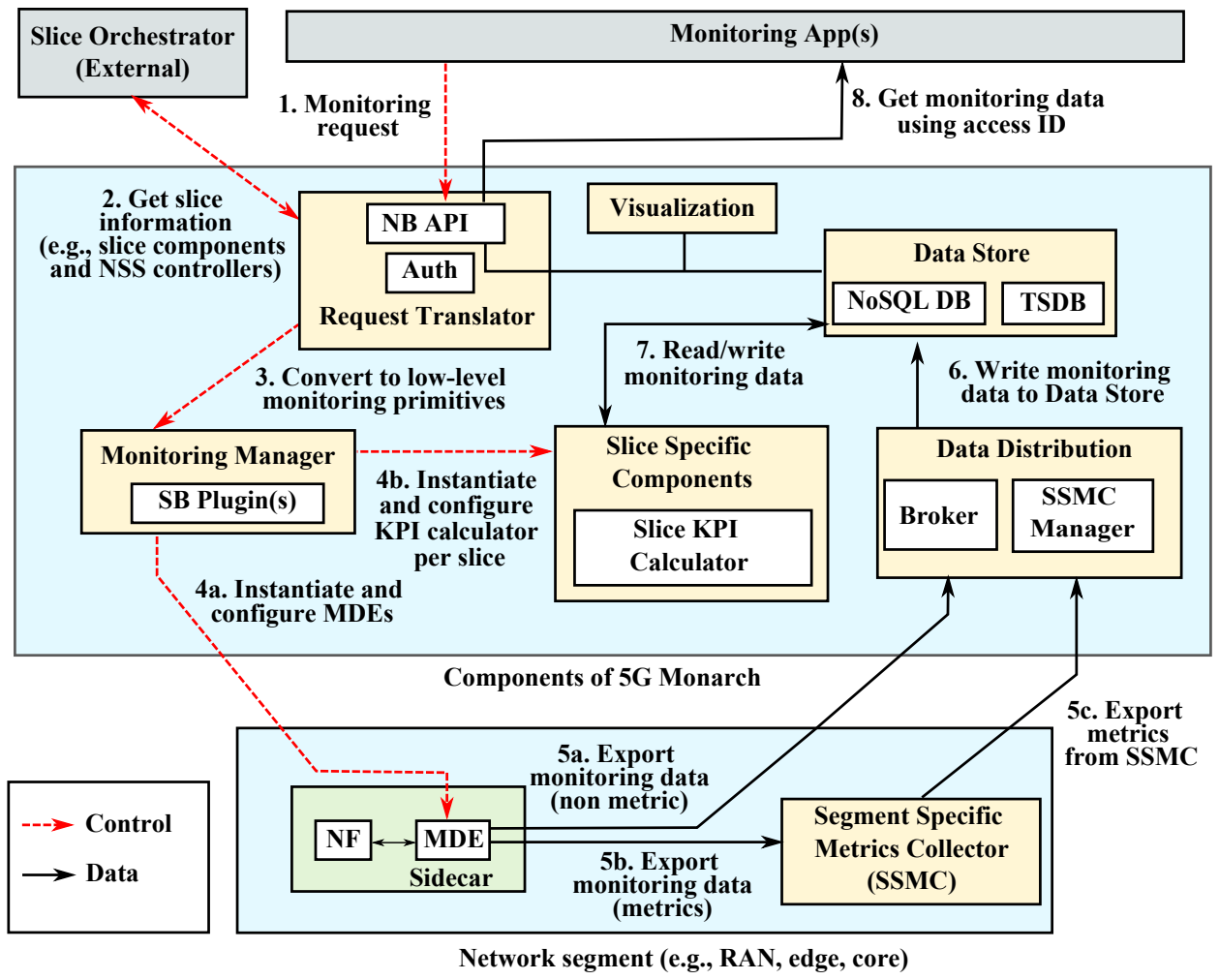

Fig. 1: Conceptual architecture for MonArch

As shown in Figure 1, the MonArch architecture comprises several high-level components, including:

- **Request translator**: The Northbound API responsible for converting high-level monitoring requests into low-level monitoring primitives and exposing computed KPIs to the user. It allows the user to specify various details related to the monitoring request, such as type, aggregation level, and more.
- **Monitoring manager**: This component is responsible for installing and configuring monitoring exporters to collect monitoring data from different NFs that make up a network slice. It uses a framework with southbound plugins that interact with the components in various network segments and provide a uniform interface to other components of the monitoring architecture. The plugin-based approach allows the use of various monitoring methods such as probing and in-band telemetry for E2E monitoring. It also communicates with an external slice orchestrator (e.g., ONAP [10]) to obtain the mapping between containerized NFs and slices.
- **Monitoring data exporter (MDE)**: MDEs are lightweight containers that are placed alongside each NF to collect monitoring information from the corresponding NF. There are two types of exporters being used: FileBeat [11] for non-numeric data such as slice to NF mapping, and NF logs, and Prometheus [12] for metric data such as network tx/rx bytes. The MDEs are responsible for extracting data from the corresponding NF, transforming the data into the correct format, and sending it to data stores. The MDEs can be written to parse and interpret vendor-specific proprietary protocols and data formatting, and convert the data into a common format used by MonArch.
- **Slice specific components**: These components are instantiated per network slice and contain the logic for network slice KPI computation. They are responsible for querying the data store, computing KPI (e.g., by correlating data from different NFs) and subsequently writing the computed KPIs back to the data store. They contain the logic necessary for computing network slice KPIs (e.g., those standardized by the 3GPP [7]) and are extendable via scripts to support custom KPIs.
- **Data store**: The data store component in MonArch is an abstraction of persistent storage and is responsible for storing monitoring and configuration data. It contains a combination of NoSQL and time-series database (TSDB). The TSDB is used to efficiently store large volumes of numeric metric data generated by monitoring different components of a 5G network over time. The NoSQL database is used to store non-numeric configuration data such as slice components and PDU sessions, as they are not well suited for TSDBs.
- **Data distribution**: The data distribution component is responsible for collecting and enriching non-metric data from various network slice segments (NSS), as well as from other sources of monitoring data instantiated by the southbound plugins (e.g., in-band telemetry collectors). This component consists of a high-performance broker (such as Apache Kafka [13]) and components to transform and enrich streaming data (e.g., Logstash [14]).
- **Segment specific metrics collector (SSMC)**: The SSMC plays a crucial role in the monitoring process by collecting metric data from various segments of the network (e.g., RAN, edge, core) and sending them to a central data store. The use of SSMCs has several advantages, including avoiding data collection across network segment boundaries, providing local persistence of monitoring data, increasing scalability through collection of aggregated metrics, and allowing for segment-wise labelling of each collected metric to aid in KPI computation.

In the subsequent section, we discuss how the different components of MonArch work together to serve a network slice monitoring request.

### B. Flow of a monitoring request

Figure 1 shows the flow of a monitoring request through 5G MonArch, which proceeds as follows:

- **Step 1**. The monitoring application specifies a high-level network slice monitoring request in JSON format, and receives an access identifier.

- **Step 2**. The request is passed to the monitoring manager, which communicates with an external slice orchestrator to gather information about the requested slice's components.
- **Step 3**. The monitoring manager translates the high-level request into low-level monitoring primitives.
- **Step 4**: The monitoring manager instantiates and configures MDEs and slice-specific KPI calculators.
- **Step 5**: MDEs export non-metric data and SSMC collects and exports metric data to the data store.
- **Step 6**: The data is written to the appropriate database in the data store.
- **Step 7**: The slice KPI calculator components perform the slice KPI computation according to predefined logic.
- **Step 8**: The monitoring data is accessed using the NB API or a data visualization module, with the access ID from step 1.

### C. *Monitoring API*

A flexible northbound API for monitoring, which provides a high-level abstraction to monitoring applications, is essential for ensuring that the apps can focus on their business logic. Some existing works, such as [6], propose to leverage the 3GPP network slice template (NST) for this purpose. However, a separate northbound API for monitoring offers greater flexibility in terms of the options that can be included, and potentially allows for the collection, aggregation, and management of monitoring data across multiple slices. Figure 2 illustrates our proposed structure for a monitoring request used by the northbound API.

```
"monitoring_request": {
    "measurement_unit": "service" | "network_function" |
     "virtualization" | "infrastructure",
    "metric": "<metric_name>",
    "entity": {
      "type": "cell" | "slice" | "session"| "user" | "...",
      "id": { "type:id": ["<list_of_ids>"] },
      "aggregation": "average" | "max" | "min" | "none"},
    "method": {
      "type": "polling" | "probing" | "in-band telemetry",
      "polling_frequency": "adaptive" | "fixed: <frequency>",
      "probing_interval": "<interval>"},
    "duration": "<duration>"}
```

Fig. 2: Structure of a monitoring request

The proposed northbound API provides a high-level abstraction to monitoring apps. The request structure includes four main components: measurement unit, which specifies the type of metric or KPI to be measured; monitoring entity, which identifies the entity to be monitored and includes three required fields (type, id, and aggregation); method, which specifies the method for collecting the monitoring data and includes a monitoring granularity field; and duration, which specifies how long the request should be active. The request structure is used to submit a high-level monitoring request to MonArch and the response includes a status field and a request id, which can be used to list, delete, or update the monitoring requests.

## IV. Case Study: Network Slice Throughput Monitoring

In this section, we discuss the implementation of MonArch in a 5G network testbed and show how it can be used to monitor *network slice throughput* KPI.

### A. *Implementation of MonArch*

The MonArch components are integrated with a 5G testbed at the University of Waterloo, in a cloud-native deployment where the 5G network and MonArch components are instantiated as containers on a Kubernetes cluster, with the MDEs instantiated as sidecar containers inside Kubernetes pods. The implementation uses various tools, which are listed in Table II. The container images, manifest files, and source code used for the implementation are publicly available on GitHub [8].

TABLE II: Implementation of MonArch components

| **Component** | **Implementation** |
|---|---|
| Monitoring manager | Kubernetes API v1.23.6 as southbound plugin |
| Data store | • NoSQL database (Elasticsearch v8.2.0)<br>• Time-series database (Prometheus v2.38.0) |
| Data distribution | • Broker (Apache Kafka v3.2.0)<br>• Enrichment (Logstash v8.2.0) |
| SSMC | Prometheus v2.38.0 |
| MDE | • Non-metric data (Filebeat v8.2.0)<br>• Metric data (Python-based Prometheus exporter) |
| KPI calculator | Python-based implementation |

The MonArch system uses Elasticsearch NoSQL database to store non-numeric configuration data and TSDB to store metric data efficiently. Data is collected from different network components by a data distribution module which is implemented using Apache Kafka for scalability. Logstash is used to clean and enrich non-metric data. The SSMC is implemented using Prometheus, which operates using a pull-based model, and allows for decoupling of MDEs from the collection system, easy identification of client error, and easy integration with cloud-native orchestration platforms like Kubernetes.

Next, we turn our attention to the monitoring API. Figure 3 shows how the monitoring API discussed in Section III-C can be used for the specific case where we want to monitor network slice throughput. Here we specify the measurement entity as *slice*, identified by its S-NSSAI, and specify a fixed polling frequency of 5s.

```
"monitoring_request":{
        "measurement_unit":"service",
        "metric":"throughput",
        "entity":{"type":"slice","id":{"snssai":["snssai"]},
                "aggregation":"none"},
        "method":{"type":"polling","polling_frequency":"fixed: 5s"},
        "duration":"10m"}}
```

Fig. 3: Example monitoring request for network slice throughput

The current case study has a limitation in the automatic conversion of high-level monitoring requests into low-level monitoring primitives, which is Step 3 of the monitoring process in Section III-B. This is an area for future work and for now, the MDE configurations are manually specified in the Kubernetes manifest files.

### B. Slice Throughput KPI computation

In order to compute network slice throughput KPI, MonArch aggregates the transmitted and received bytes for every PDU session associated with each UPF instance in the slice. This requires correlating information from both the SMF and UPF. To achieve this, MonArch instantiates two MDEs alongside the SMF and UPF, using the Kubernetes API. The MDE for SMF collects information about the mapping between slices and their active PDU sessions, while the MDE for UPF exports information about the bytes transmitted and received for each PDU session, properly labeled with information such as the originating UPF instance, direction (uplink/downlink), and network segment.

The slice-specific KPI computation module then queries the data store to collect information about the slice and the number of bytes received and transmitted per active PDU session. It filters the list of PDUs for a given slice using S-NSSAI to PDU session mapping and aggregates the received and transmitted bytes for these filtered PDU sessions to calculate slice throughput. MonArch's ability to label metrics by direction and per PDU session provides flexibility for monitoring applications.

## V. Evaluation

### A. Experimental Setup

**Testbed**: We evaluate the performance of the proposed MonArch architecture on a 5G testbed established at the University of Waterloo, which is described in detail in [15]. The testbed comprises of a 5G mobile core based on Free5GC, emulated gNodeB and UE using UERANSIM, and all network functions are run as containers on a 6 node Kubernetes cluster, consisting of Intel NUC PCs (Intel i7-6770HQ (4) @ 2.600 GHz with 16GB RAM). The cluster also runs Prometheus as the SSMC. The other components of the MonArch architecture, such as data store, are hosted on a separate 4 node Kubernetes cluster, which are powered by Intel Xeon servers (E3-1230 v3 (8) @ 3.7GHz) with 16GB RAM.

**Network slicing scenario**: To evaluate the performance of the MonArch architecture, we created three network slices on the testbed, each with two UE initiated PDU sessions for a total of six PDU sessions. Each slice has its own dedicated instances of SMF and UPF, while all other 5G core functions are common across all slices. We used the case study described in Section IV to collect data for the slice throughput KPI computation. The results for different scenarios were calculated by repeating each scenario 10 times and presenting the average results.

### B. Performance Metrics

**Resource usage**: The CPU and memory usage of the SSMC. To calculate CPU usage, we leverage CPU metrics exposed by cAdvisor integrated with Kubernetes.

**Monitoring overhead**: The amount of network traffic (bytes/sec) transmitted by the MDEs.

**Ingestion time**: The time taken by the SSMC to fetch data from the MDEs, given by Prometheus *scrape_duration* metric.

### C. Results

In order to assess the performance of MonArch, we conducted evaluations to measure the impact of the number of slices and the monitoring interval on the system's resource usage. Specifically, we collected non-metric data (slice information) during the creation of slices, and periodically collected metric data from the SSMC at the specified monitoring interval. The focus of our evaluation was primarily on the resource usage of the SSMC.

**Impact of number of slices**: As shown in Figure 4a, the CPU usage of the SSMC increases linearly with an increase in the number of PDU sessions, while the monitoring interval is fixed at 5 seconds. Each slice has 2 PDU sessions, therefore an increase in PDU sessions from 2 to 3 represents the addition of a second network slice. The results indicate that the increase in CPU consumption of the SSMC is reasonable, with a 3x increase in the number of slices resulting in a 1.4x increase in CPU usage.

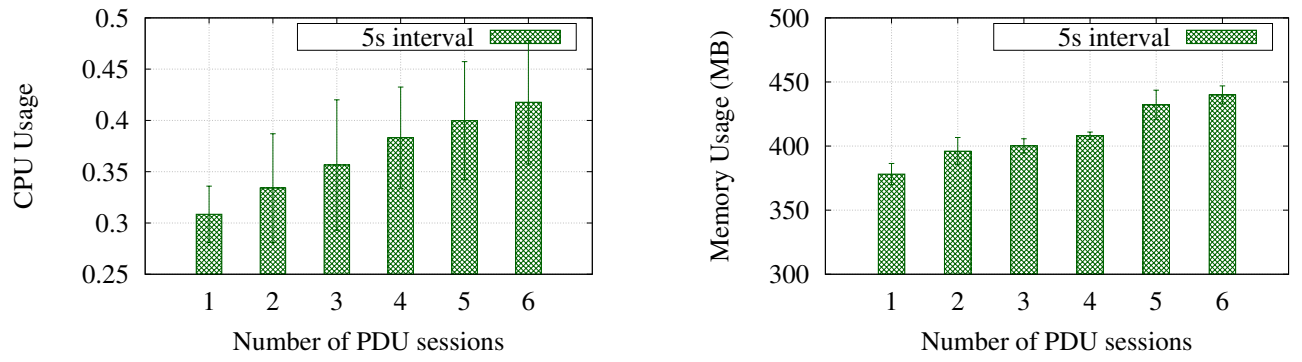

(a) CPU consumption (b) Memory usage

Fig. 4: SSMC resource usage with varying slices

Figure 4b shows that a 3x increase in the number of slices results in approximately a 1.1x increase in the average memory usage of the SSMC. Despite the increases, the values of CPU (40%) and memory consumption (450MB) for 3 slices are considered reasonable, indicating that the SSMC can handle a larger number of slices.

Our approach of using a single SSMC to collect metrics from all slices, as opposed to using slice-specific collectors, allows for a more scalable monitoring architecture. The resource usage of the SSMC, as shown in Figure 4, increases linearly with the number of slices, but at a reasonable rate. Additionally, by having the data transformation complexity handled by NF-specific MDEs, the architecture is more flexible and changes to a specific NF will not affect other parts of the monitoring system. This approach is in contrast to previous work such as [6] which had slice-specific collectors.

**Impact of varying monitoring interval**: The performance of MonArch is impacted by the monitoring interval, as it determines the frequency of data collection from the MDEs. Our analysis of the SSMC CPU usage and memory usage, captured in Figures 5a and 5b respectively, shows that as the monitoring interval increases, the CPU usage and memory usage decrease. The decrease in CPU usage is significant, with a reduction of approximately 60% when changing the monitoring interval from 1 second to 3 seconds. We also observe that the decrease in CPU usage gradually tapers off as the monitoring interval increases further. Furthermore, we see that as the monitoring interval decreases, more data points need to be kept in RAM, leading to an increase in memory usage, which is reflected in the Figure 5b.

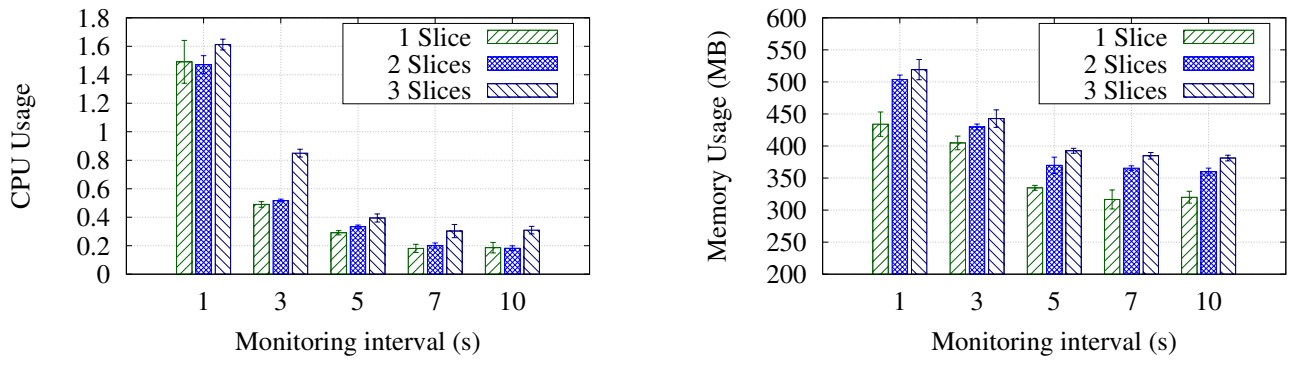

(a) CPU consumption (b) Memory usage

Fig. 5: SSMC resource usage with varying monitoring interval

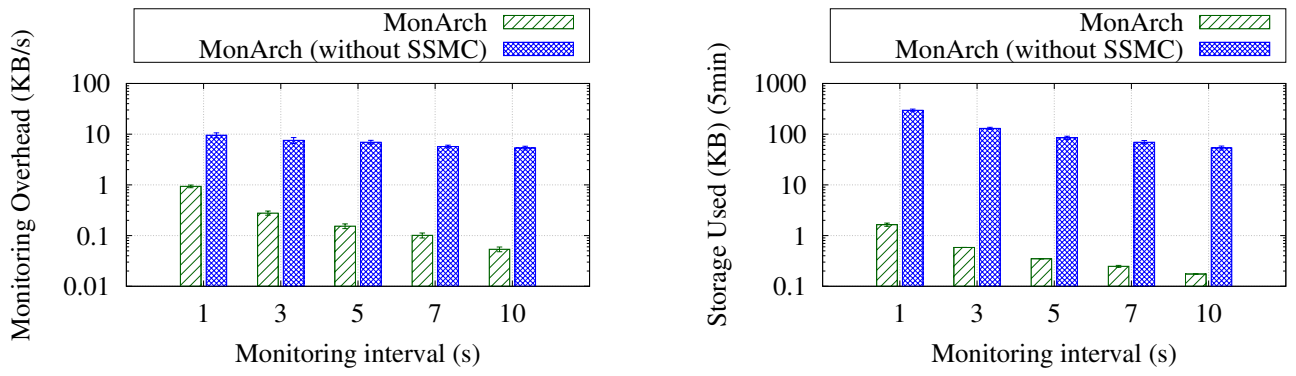

(a) Monitoring overhead (b) Storage used

Fig. 6: Comparison of MonArch with and without SSMC

In MonArch, two separate methods are used for collecting data, depending on whether it is numeric (metric) or non-numeric. Numeric data is sent to a TSDB through the SSMC for efficient collection and storage. This architectural choice is compared to a variant of MonArch (without SSMC) in terms of monitoring overhead and storage usage, as shown in Figure 6. Figure 6a shows the monitoring overhead of the UPF MDE at different monitoring intervals. A notable decrease (approximately 3x) in the monitoring overhead is observed on increasing monitoring interval from 1s to 3s. As the monitoring interval is increased further, this decrease tapers off. Additionally, there is an almost 10x decrease in the monitoring overhead compared to the MonArch variant without SSMC. This is due to the efficient storage of metric data in a compressed format in a TSDB, as opposed to storing them in a NoSQL database. Similarly, Figure 6b shows a significant reduction of storage overhead achieved by MonArch compared to its variant without SSMC.

Next, we examine the ingestion time of MonArch, as illustrated in Figure 7a. Our findings reveal that there is minimal impact on the SSMC ingestion time, regardless of the monitoring interval or the number of slices. This suggests that MonArch is able to scale effectively without introducing additional delays. However, we also find that while the monitoring interval does not significantly affect the ingestion time, it has a significant impact on the accuracy of the monitoring results. To evaluate the impact of the monitoring interval on accuracy, we conduct an experiment where we send time-varying traffic through a network slice, measure the slice throughput with MonArch and compare the results to the ground truth. Figure 7b illustrates that accuracy decreases as the monitoring interval increases from 1s to 10s, as indicated by the deviation from the ground truth during sharp changes in throughput, such as at the 25s and 30s mark. Based on the results of Figures 6 and 7b, we conclude that a monitoring interval of 5 seconds offers a reasonable tradeoff between monitoring overhead and accuracy.

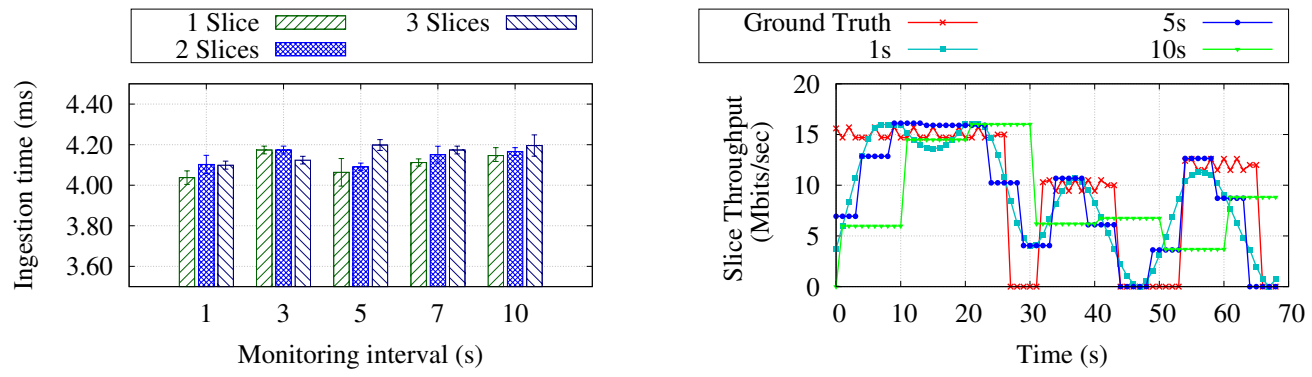

(a) Data ingestion time (b) Slice throughput

Fig. 7: Impact of varying monitoring interval

**Insights from MonArch**: MonArch is a scalable monitoring architecture for cloud-native 5G deployments that allows for the collection and computation of a wide range of network slice metrics. The architecture uses Prometheus as the SSMC, which allows for easy collection of metrics in cloud-native environments and supports a wide range of instrumentation. The architecture can be easily extended to compute other network slice KPIs, such as slice connection density, and can be further enhanced by incorporating specialized data collection mechanisms, such as in-band telemetry (INT) for high-resolution monitoring of the network data plane. The current version of MonArch uses a central TSDB to collect data from the SSMCs in each network segment, which may limit scalability. However, this can be addressed by implementing the TSDB using a high-availability Prometheus alternative like Thanos, which allows for fast querying of the SSMC on-demand.

## VI. Conclusion

In this paper, we present MonArch, a scalable monitoring architecture for cloud-native 5G deployments that focuses on network slice monitoring and computation of network slice KPIs. We validate MonArch by implementing it on a 5G testbed and using it to compute network slice throughput KPI. Our evaluation results show that MonArch scales well with the number of slices and that a monitoring interval of 5 seconds offers a good balance between monitoring overhead and accuracy. In future work, we plan to implement automatic translation of high-level monitoring requests and incorporate In-Band Telemetry (INT) as a plugin component for improved E2E monitoring.

## Acknowledgements

This work was supported in part by Rogers Communications Canada Inc. and in part by a Mitacs Accelerate Grant.